\documentclass[review]{elsarticle}

\usepackage[margin=1in]{geometry}
\usepackage{amssymb}
\usepackage{amsmath}
\usepackage{multirow}

  \biboptions{authoryear,colon,round}

\begin{document}

\begin{frontmatter}




\title{Cubic Single Crystal Representations in Classical and Size-dependent Couple Stress Elasticity}


\author[label1]{D. Bansal}
\author[label2]{G.F. Dargush\corref{cor1}}
\ead{gdargush@buffalo.edu}
\cortext[cor1]{Corresponding author. Tel.: +1 716 645 2315; fax: +1 716 645 2883.}
\author[label1]{A.J. Aref}
\author[label2]{A.R. Hadjesfandiari}

\address[label1]{Department of Civil, Structural, and Environmental Engineering, University at Buffalo, New York, 14260, USA}
\address[label2]{Department of Mechanical and Aerospace Engineering, University at Buffalo, New York, 14260, USA}


\begin{abstract}
Beginning with Cosserat theory in the early 20th century, there have been several different formulations for size-dependent elastic response.
In this paper, we concentrate on the application of classical Cauchy theory and the recent parsimonious consistent couple stress theory to model a homogeneous linear elastic solid, exemplified by a pure single crystal with cubic structure.  The focus is on an examination of elastodynamic response based upon wave velocities from ultrasonic excitation and phonon dispersion curves, along with adiabatic bulk moduli measurements.  In particular, we consider in detail elastic parameter estimation within classical elasticity and consistent couple stress theory for four different cubic single crystals (NaCl, KCl, Cu, CuZn).  The classical theory requires the estimation of three independent material parameters, while only one additional parameter relating skew-symmetric mean curvature to skew-symmetric couple-stress is needed for the size-dependent consistent couple stress theory.  This additional parameter can be defined for cubic crystals in terms of a material length scale, which is found to be on the order of tens of microns for the four materials studied here.  Furthermore, a detailed statistical investigation provides strong to very strong evidence that couple stress theory is superior to classical Cauchy elasticity for representing the wave velocities and adiabatic bulk moduli for all four single crystals.  
\end{abstract}
\vspace{12pt}

\end{frontmatter}




\section{Introduction}
The theory of elasticity has played an essential role in the mechanics of solid continua since 1822, when it was first introduced by Cauchy to the Academy of Sciences in Paris. The initial version, conceptualized from a model of point particles interacting through central forces, relied on symmetry of stress and strain tensors with a single elastic parameter for isotropic materials.  However, physical experiments soon showed that more generality was needed and a revised version was developed that incorporated a second independent material parameter for the isotropic case and a total of three independent material parameters for cubic crystals.  This latter version, which retains the Cauchy assumption to ignore the possibility of couple-tractions and body-couples, remains intact to this day as the classical theory of elasticity.

Thus, classical Cauchy elasticity stands on the idea that angular momentum of the force-tractions about principal directions or coordinate axes when summed up should go to zero.  However, as argued first by \cite{Voigt1887}, then by Raman in a series of papers from his research group in the mid-20th century, the tractive forces must be considered over finite volume elements, rather than over volumes so small that can be regarded as point particles \citep{Viswa1955, Raman-Krishna9-1955, Raman-Viswa8-1955},  and therefore, further consideration must be given to the balance of angular momentum.  Similar arguments are given by proponents of micropolar theory \citep{Cosserat1909, Mindlin1964, Eringen1968, Nowacki1986}, couple stress theory \citep{Mindlin1962, Koiter1964, Ali2011} and second gradient theories \citep{AltanAifantes1997, Yang2002, Lazar2005} for the response of solids at smaller scales.  

There have been many attempts to present a generalized non-classical theory of elasticity in a consistent form.  Initial work in the area, in addition to that cited in the previous paragraph, includes the theoretical development of Cosserat (micropolar) elasticity, void elasticity \citep{Cowin1983}, second gradient theories, nonlocal elasticity \citep{Eringen1972} and micromorphic elasticity \citep{Mindlin1965, Eringen1968-1}.  Brief descriptions with advantages and limitations of each of these theories are provided in a review paper by \cite{Lakes1995}.  Many of the applications of these theories have been to composite materials, where the situation is complicated by all of the interactions between the dissimilar constituents.  Naturally, along with the ever increasing push to develop new materials and devices at the micro- and nano-level comes the critical need to formulate self-consistent theories to capture size-dependent mechanical response at the finest scales for which a continuum representation is valid.  

A recently developed theory, based upon skew-symmetric couple-stress and mean curvature tensors, has been shown to be fully self-consistent \citep{Ali2011}.  With the reduction in stress components due to this skew-symmetry, all indeterminacies are eliminated.  Furthermore, the theory is parsimonious in terms of material parameters, having only a single additional couple stress parameter for isotropic and cubic materials.  However, questions remain as to its relevance to real, physical solids.  Is the material length scale, herein defined as $l$, large enough to permit continuum representations at or below that scale?  If, for example, $l$ is on the order of the atomic spacing, then couple stress theory would be merely of theoretical interest, because the theory would only predict significant deviations from classical Cauchy theory at length scales for which a continuum representation is not appropriate.  A second and yet more fundamental question is whether actual materials follow tensorial mechanics below the classical continuum range.  Only by comparing theoretical predictions with the results of physical experiments can this begin to be answered.  Thus, the primary motivation of the present paper is to address these two foundational questions by examining the dynamic response of cubic single crystals, which represent solids in perhaps their purist form for study as a continuum.

With our focus on single crystals, where the internal structure is well-known, we attempt to answer the following more specific questions.  How well does the classical theory represent the response of cubic crystals, especially at smaller scales?  Are three parameters sufficient to capture the elastic response?  Is the Cauchy assumption mentioned above, which leads to symmetry of the force-stress tensor, universally valid?  If the classical theory is found lacking, is consistent couple stress theory better supported by the experimental data?  We begin in the following section by reviewing several non-classical elasticity theories.

%
%

\section{Review of Non-classical Elastodynamic Theories}
The most prominent of the size-dependent theories have been Cosserat elasticity \citep{Cosserat1909} and its various direct offshoots, including micropolar \citep{Eringen1968} and the more general micromorphic \citep{Eringen1968-1} theories.  These theories attempt to represent the effect of the discontinuous microstructure of materials by introducing additional degrees of freedom as continuous microfields 
independent from the macroscopic displacements.  The concept of independent rotations originated from beam and plate theories of structural mechanics, in which one- and two-dimensional objects are embedded in higher dimensional spaces.  
In micromorphic elasticity, infinitesimal elements of matter at each point in the continuum representation of the solid can translate, rotate and deform microscopically, and 18 elastic constants are required for the isotropic case \citep{Mindlin1965, Eringen1968-1}.  The theory predicts dispersion for both dilatational and shear waves.  Cosserat elasticity has been shown to be a special case of micromorphic elasticity \citep{Lakes1995}, but still requires six independent coefficients for isotropic elasticity.

In void theory, the change of volume fraction has been taken as a kinematical variable, while no consideration has been given to rotation. 
Void theory gives rise to two types of dilatational waves and one shear wave. For dilatational waves, one is similar to the dilatational wave of classical elasticity and the other wave carries a change in the void volume fraction, while the shear waves propagate in the medium without dispersion \citep{Puri-Cowin1985, Lakes1995}, thus giving three different wave speeds even at the bulk level in the \{100\} direction. 


Let us turn then to couple stress theories, in their various forms, as potential candidates for representing the size-dependent elastodynamic response of homogeneous solids, such as cubic single crystals, with a minimal number of additional material parameters.  The kinematical quantities in the original couple stress theory are displacements and macrorotation, representing half of the curl of the displacement field \citep{Mindlin1962, Koiter1964}. The gradient of the rotation vector (curvature tensor) is then used in the formulation of the stress-strain relationships. However, these developments have suffered from indeterminacy of the spherical part of the couple-stress tensor and the appearance of the body-couple in the relation for the force-stress tensor, leading to the designatation of the theory as ``indeterminate couple stress theory" \citep{Eringen1968, Ali2011}.  One well-cited attempt to eliminate indeterminacy is the modified couple stress theory of \cite{Yang2002}, which unfortunately requires an unsubstantiated moment of angular momentum balance law to impose symmetry of the couple-stress tensor.  
In light of the above considerations, we find that the inclusion of size-dependence of the material in continuum mechanics theory has been a difficult challenge.

\begin{figure}
\begin{center}
\includegraphics[width=0.75\textwidth]{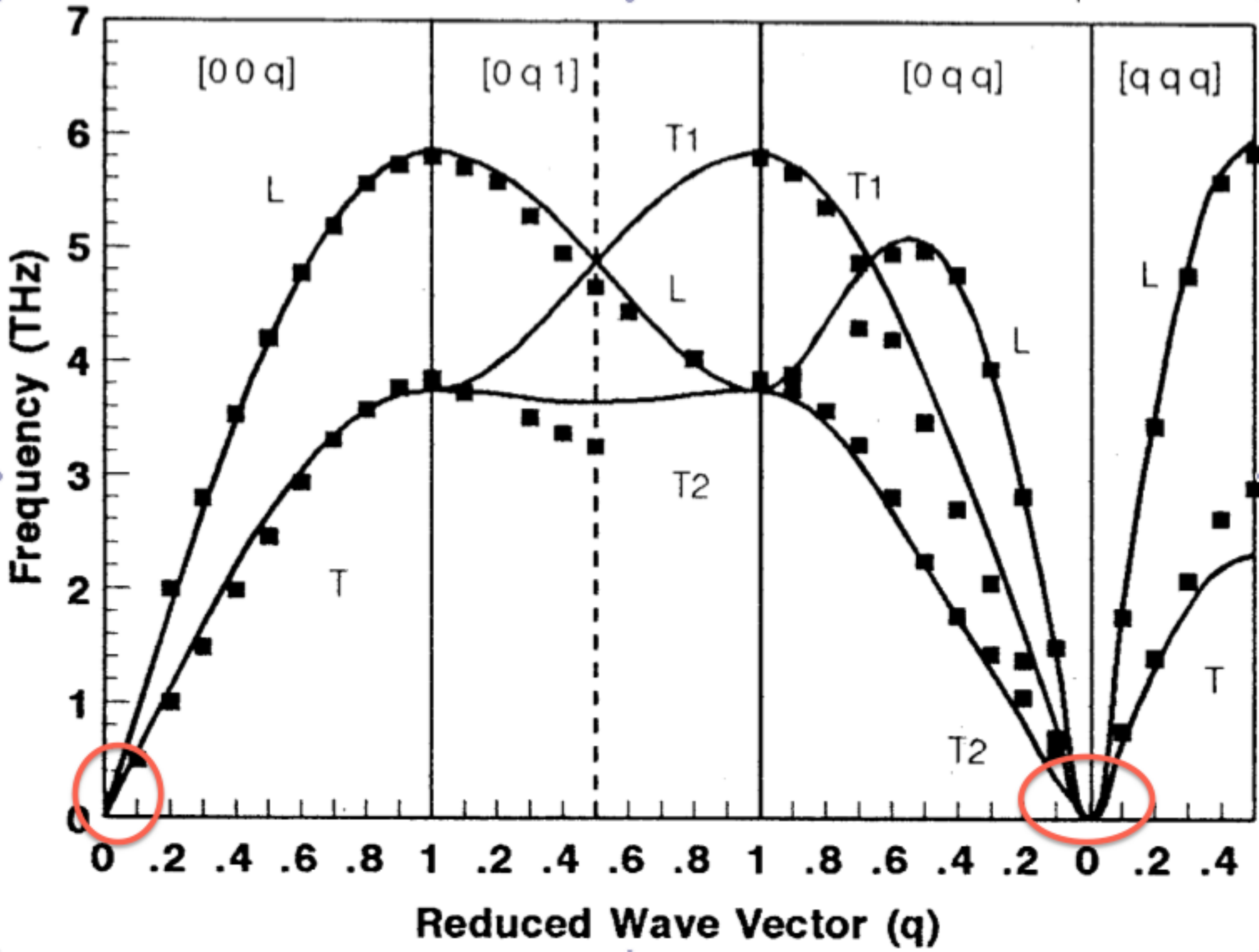}
\caption{\label{P-PD}Phonon dispersion relation for platinum showing the experimentally unexplored region (adapted from \cite{Platinum}).}
\end{center}
\end{figure}

Figure~\ref{P-PD} displays the phonon dispersion relations in terms of frequency versus reduced wave vector for platinum along several directions.  As indicated by the circled regions in the figure, wave dispersion in the frequency range from gigahertz to roughly 0.5 terahertz very close to the zone center is largely unknown, primarily because of the challenge to measure the wave velocity in this region with inelastic neutron or x-ray scattering.  Classical continuum mechanics predicts no dispersion in this region, leading to the same value of elastic constants with frequency or wavelength, while most non-classical theories will exhibit scale dependence in this range.  Interestingly, some researchers have tried to explain the dispersion relations for the terahertz plus frequency range using non-classical continuum theories.  Examples include work in strain gradient elasticity \citep{DiVincenzo1986, AltanAifantes1997, Every2005, Everyetal2006, Maranganti2007, Maranganti2007jmps, AskesAifantes2011, Shodjaetal2013} and in micromorphic theory \citep{Youping1, Youping2}.
On the other hand, it becomes abundantly clear that we need more data in the circled unexplored region of the phonon dispersion graphs, because this is precisely the region where many technological advances are taking place.  All micro and nano level mechanics, which often has been difficult to explain theoretically, is hidden in this unexplored area.

In recent work, the indeterminacy in couple stress theory has been resolved by establishing the skew-symmetric character of the couple-stress tensor in size-dependent continuum representations of matter \citep{Ali2011, Ali-Gary2015}. This consistent theory also satisfies the admissible boundary conditions and the principle of virtual work.  
In addition, couple stress theory specifically deals with the unexplored region in Figure 1 and may explain the dispersion relations or size-dependency in the material below the scale, where we can adequately represent the material with classical continuum mechanics.  Several aspects of wave propagation under size-dependent theory for isotropic thermoelastic media are discussed in recent work \citep{Ali2014thermo}.  However, because of the unavailability of experimental data, at this point in time, it is hard to identify the cut-off frequency or wavelength for which couple stress theory is no longer valid and must be replaced by an atomistic representation. Another interesting point to note is that in the formulation of consistent couple stress theory, no consideration has been given to wave propagation, yet when we solve for wave propagation in cubic crystals, we find results in accordance with experimental evidence.  For example, for cubic single crystals, in the \{100\} direction, two transverse waves overlap with each other, while in the \{110\} direction, all three waves have different dispersion relations.

Within this paper, we adopt consistent couple stress theory to solve for wave propagation in the most general anisotropic material having 45 constants, and further determine the four elastic constants (in contrast to three for Cauchy elasticity) for simple cubic (KCl, NaCl), face-centered cubic (Cu) and body-centered cubic (CuZn) single crystals.  For cubic materials, wave dispersion has been observed in only transverse directions. As explained by \cite{Raman-Krishna9-1955} in their memoir, material behavior of cubic crystal at high frequencies/smaller scale cannot be explained through just three elastic constants $A_{11}$, $A_{12}$ and $A_{44}$.  They showed by means of calculating the bulk modulus that if we only consider three elastic constants for a cubic crystal, the bulk modulus of the material differs significantly from the experimental measurements.  This suggests that we need an additional constant to model the elastic material behavior, which is the problem we begin to address in the next section. 


\section{Wave Propagation in Isotropic and Anisotropic Media} \label{waveprop}
The governing differential equations for linear and angular momentum balance under consistent couple stress theory for infinitesimal deformations can be written, respectively, as \citep{Ali2011, Ali2014thermo}:
\begin{subequations}
\label{cs-pde}
\begin{align}
\sigma_{ji,j}+F_i&=\rho \ddot{u_i}\label{pdefirst} \\
\mu_{ji,j}+\varepsilon_{ijk} \sigma_{jk}&=0\label{pdesecond}
\end{align}
\end{subequations}
where standard indicial notation is employed with summation over repeated indices, an index following a comma denoting a spatial derivative and a superposed dot representing partial differentiation with respect to time.  In Eq.~\eqref{cs-pde}, $\sigma_{ij}$ and $\mu_{ij}$ are the force-stress and couple-stress tensors, respectively, $u_i$ represents the displacement field, $F_i$ is the body-force per unit volume, $\rho$ is the mass density of the material and $\varepsilon_{ijk}$ denotes the permutation or Levi-Civita symbol.  Within this theory, $\sigma_{ij}$ is a general, true (polar) non-symmetric tensor, while $\mu_{ij}$ is a pseudo (axial) skew-symmetric tensor, which can be written instead in terms of its dual (polar) vector $\mu_i=\varepsilon_{ijk}\mu_{kj}/2$.  

Constitutive relations are needed to close the set.  For an isotropic linear elastic material, one finds the following set \citep{Ali2011}:
\begin{subequations}
\label{cs-ceiso}
\begin{align}
\sigma_{(ij)} &= 2\mu e_{ij} + \lambda e_{kk} \delta_{ij}\label{isofirst} \\
\mu_{i} &= -8 \eta \kappa_i  \label{isosecond}
\end{align}
\end{subequations}
with parentheses around indices to repesent the symmetric part of a second order tensor, $e_{ij}$ as the usual infinitesimal strain tensor, $\kappa_i$ as a polar mean curvature vector, $\delta_{ij}$ denoting the Kronecker delta, scalars $\mu$ and $\lambda$ as the classical Lam\'{e} elastic coefficients and with $\eta$ as the single additional elastic material coefficient in this consistent couple stress theory.  Thus, Eq. \eqref{isofirst} relates force-stresses to strains and is associated with the storage of strain energy, as in the classical theory.  Equation \eqref{isosecond} is the result of an additional storage mechanism, involving curvature energy, providing the link between couple-stresses and mean curvatures.  We should emphasize that both kinematic measures of deformation are related directly to the displacement field through the relations:
\begin{subequations}
\label{cs-kin}
\begin{align}
e_{ij} &= u_{(i,j)}\label{kinfirst} \\
\kappa_i &= (u_{k,ki}-u_{i,kk})/4 \label{kinsecond}
\end{align}
\end{subequations}
Furthermore, for the isotropic case, we may write $\eta=\mu\,l^2$, with $l$ as a length scale.  Typically, the relative contribution of curvature energy will become important as the characteristic length of an elastic problem approaches that scale.

By substituting Eqs.~\eqref{cs-kin} into Eqs.~\eqref{cs-ceiso} and then into Eq.~\eqref{pdesecond}, we may write the following relation for the force-stress:
\begin{align}
\sigma_{ij} &= 2\mu e_{ij} + \lambda e_{kk} + 2\eta \nabla^2\omega_{ij}\label{forcestress}
\end{align}
in terms of the Laplacian of the rotation tensor, which in turn is defined as the skew-symmetric part of the gradient of displacement.   Thus,
\begin{align}
\omega_{ij} &= u_{[i,j]}\label{rotation}
\end{align}
with square brackets around indices signifying the skew-symmetric part of a second order tensor.  Notice from Eq.~\eqref{forcestress} that consistent couple stress theory is not a strain gradient formulation, but rather depends upon displacements and rotations as fundamental variables.

An extended Navier equation for a uniform elastic isotropic body under consistent couple stress theory also may be written in the following form by manipulating the above relations \citep{Ali2011, Ali2014thermo}:
\begin{align}
(\lambda+2\mu)u_{j,ij}+(\mu-\eta\nabla^2)(\nabla^2 u_i-u_{j,ij}) +F_i &=\rho \ddot u_i\label{navierplus}
\end{align}
This form suggests decomposition of the response into irrotational and solenoidal parts, exactly as in the classical case.  As we shall see below, the irrotational or P-wave behaves identically to its classical counterpart, whereas the solenoidal or S-waves are dispersive in consistent couple stress theory due to the influence of the fourth order derivatives appearing in Eq.~\eqref{navierplus} associated with curvature energy.

The situation is naturally much more complicated for general anisotropic material response.  The constitutive relations in this case can be written in tensorial form as follows \citep{Ali2011}:
\begin{subequations}
\label{cs-ceaniso}
\begin{align}
\sigma_{(ij)} &= A_{ijkl} e_{kl} + C_{ijk} \kappa_k\label{anisofirst} \\
\mu_{i} &= - {1\over 2} C_{jki} e_{jk}  - {1\over 2} B_{ij} \kappa_j  \label{anisosecond}
\end{align}
\end{subequations}
where $A_{ijkl}$, $C_{ijk}$, and $B_{ij}$ are the fourth, third, and second order constitutive tensor coefficients, respectively.  For the most general linear anisotropic material, $A_{ijkl}$, $C_{ijk}$, and $B_{ij}$ have 21, 18, and 6 distinct components, respectively.  For centrosymmetric materials, all components of the $C_{ijk}$ tensor vanish, which means that there is no coupling between strain and curvature energies at the constitutive level.  More generally, substituting Eqs.~\eqref{cs-ceaniso} into the governing balance laws Eqs.~\eqref{cs-pde} for a uniform anisotropic solid within skew-symmetric couple stress theory produces the following size-dependent equations of motion in displacement form \citep{Ali2011, Ali2014thermo}:
\begin{align}
A_{ijkl}&u_{k,lj} + \frac{1}{4}C_{ijk}\left(u_{m,mjk} - \nabla^2u_{k,j}\right) + \frac{1}{4}C_{kmi}\nabla^2u_{k,m} \nonumber \\
& - \frac{1}{4}C_{kmj}u_{k,mij} + \frac{1}{16}B_{ik}\left(\nabla^2u_{m,mk} - \nabla^2\nabla^2u_k\right) \nonumber \\
&- \frac{1}{16}B_{jk}\left(u_{m,mkij} - \nabla^2u_{k,ij}\right) + F_i = \rho \ddot{u}_{i}\label{C-S1}
\end{align}


The presence of third and fourth order spatial derivatives in \eqref{C-S1} suggests that wave propagation may be dispersive.  To investigate this phenomenon, let us next assume a plane wave solution of the form
\begin{align}\label{C-S2}
u_n = A_0p_n\exp\{i\omega t\}\exp\{-ikd_jx_j\}
\end{align}
where $A_0$ is the amplitude of the wave, $p_n$ is the polarization vector, 
$\omega$ is angular frequency, $\textbf{k} = \textbf{d}k$ is the wave vector, $\bf{d}$ is the propagation unit vector, $k = \sqrt{k_1^2 + k_2^2 + k_3^2}$ is the wave number and $i$ is the unit imaginary number. Using Eq.~\eqref{C-S2}, the following can be written
\begin{align}
u_{k,l} &= (-ikd_j\delta_{jl})u_k \nonumber \\
u_{k,lj} &= (-k^2d_ld_j)u_k \nonumber \\
u_{m,mjk} &= (ik^3d_md_jd_k)u_m \nonumber \\
\nabla^2u_{k,j} &= (ik^3d_jd_ld_l)u_k \nonumber \\
\nabla^2u_{k,m} &= (ik^3d_md_ld_l)u_k \nonumber \\
u_{k,mij} &= (ik^3d_md_id_j)u_k \nonumber \\
\nabla^2u_{m,mk} &= (d_md_kd_ld_l)u_m \nonumber \\
\nabla^2\nabla^2u_k &= (k^4d_md_md_ld_l)u_k \nonumber \\
u_{m,mkij} &= (k^4d_md_kd_id_j)u_m \nonumber \\
\nabla^2u_{k,ij} &= (k^4d_id_jd_ld_l)u_k\label{C-S3}
\end{align}
Substituting Eq.~\eqref{C-S3} in Eq.~\eqref{C-S1} and ignoring the body-force $F_i$, we obtain
\begin{align}
-A_{ijkl}&k^2d_ld_ju_k + \frac{1}{4}C_{ijk}\left(ik^3d_md_jd_ku_m - ik^3d_jd_ld_lu_k\right) \nonumber \\
&+ \frac{1}{4}C_{kmi}\left(ik^3d_md_ld_lu_k\right) - \frac{1}{4}C_{kmj}\left(ik^3d_md_id_ju_k\right) \nonumber \\
&+ \frac{1}{16}B_{ik}\left(k^4d_md_kd_ld_lu_m - k^4d_md_md_ld_lu_k\right) \nonumber \\ 
&- \frac{1}{16}B_{jk}\left(k^4d_md_kd_id_ju_m - k^4d_id_jd_ld_lu_k\right) \nonumber \\
&  + \rho\omega^2u_i = 0 \label{C-S4}
\end{align}
Simplification leads to
\begin{align}
&\left[-A_{ijkl}(\delta_{kn}d_ld_jk^2) + \frac{1}{16}B_{ik}(d_md_k\delta_{mn}k^4)\right. \nonumber \\
& \hspace{0.3in}\left.- \frac{1}{16}B_{ik}\delta_{kn}k^4 - \frac{1}{16}B_{jk}(d_md_kd_id_j\delta_{mn}k^4)\right. \nonumber \\ 
&\hspace{0.3in}\left.+ \frac{1}{16}B_{jk}(d_id_j\delta_{kn}k^4) + i\frac{k^3}{4}\bigg\{C_{ijk}(d_md_jd_k\right.\delta_{mn}) \nonumber \\
&\hspace{0.3in}\left. - C_{ijk}(d_j\delta_{kn}) - C_{kmj}(d_md_id_j\delta_{kn})\right. \nonumber \\
&\hspace{0.3in} + C_{kmi}(d_m\delta_{kn})\bigg\}\bigg]u_n + \rho\omega^2\delta_{in}u_n = 0 \label{C-S5}
\end{align}
In matrix notation, the first term in Eq.~\eqref{C-S5}, $A_{ijkl}(\delta_{kn}d_ld_j)$, can be written as
\begin{align}\label{C-S6}
&\mathrm{A} = A_{ijkl}\delta_{kn}d_ld_j = \nonumber\\
& \left[\begin{array}{ccc} \vspace{5pt} \bigg\{[A_{1-1-}]\{\textbf{d}\}\bigg\}^T\bigg\{\textbf{d}\bigg\} & \bigg\{[A_{1-2-}]\{\textbf{d}\}\bigg\}^T\bigg\{\textbf{d}\bigg\} & \bigg\{[A_{1-3-}]\{\textbf{d}\}\bigg\}^T\bigg\{\textbf{d}\bigg\} \\ \vspace{5pt} 
\bigg\{[A_{2-1-}]\{\textbf{d}\}\bigg\}^T\bigg\{\textbf{d}\bigg\} & \bigg\{[A_{2-2-}]\{\textbf{d}\}\bigg\}^T\bigg\{\textbf{d}\bigg\} & \bigg\{[A_{2-3-}]\{\textbf{d}\}\bigg\}^T\bigg\{\textbf{d}\bigg\} \\
\bigg\{[A_{3-1-}]\{\textbf{d}\}\bigg\}^T\bigg\{\textbf{d}\bigg\} & \bigg\{[A_{3-2-}]\{\textbf{d}\}\bigg\}^T\bigg\{\textbf{d}\bigg\} & \bigg\{[A_{3-3-}]\{\textbf{d}\}\bigg\}^T\bigg\{\textbf{d}\bigg\}
\end{array}\right]_{3\times3}
\end{align}
where 
\begin{subequations}
\label{C-S7}
\begin{align}
A_{1-1-} = \left[\begin{array}{ccc}
A_{1111} & A_{1112} & A_{1113} \\
A_{1211} & A_{1212} & A_{1213} \\
A_{1311} & A_{1312} & A_{1313}
\end{array}\right]_{3\times3}
\end{align}
\begin{align}
A_{3-2-} = \left[\begin{array}{ccc}
A_{3121} & A_{3122} & A_{3123} \\
A_{3221} & A_{3222} & A_{3223} \\
A_{3321} & A_{3322} & A_{3323}
\end{array}\right]_{3\times3}
\end{align}
\end{subequations}
and, similarly other matrices can be expanded. Further, $\textbf{d} = [d_1, d_2, d_3]^T$, and $\{\}^T$ is the transpose of the matrix. Other terms also can be expressed in matrix notation as follows:
\begin{subequations}
\label{C-S8}
\begin{align}
\mathrm{\bar B_1} = B_{ik}d_md_k\delta_{mn} = \bigg\{\big[B\big]\big\{d\big\}\bigg\}\bigg\{d\bigg\}^T \label{C-S8-1}
\end{align}
\begin{align}
\mathrm{\bar B_2} &= B_{ik}\delta_{kn} = [B] \label{C-S8-2}
\end{align}
\begin{align}
\mathrm{\bar B_3} &= B_{jk}d_md_kd_id_j\delta_{mn} \nonumber \\
&= \Bigg(\bigg\{[B]\{d\}\bigg\}^T\bigg\{d\bigg\}\Bigg)\Bigg\{d\Bigg\}\Bigg\{d\Bigg\}^T  \label{C-S8-3}
\end{align}
\begin{align}
\mathrm{\bar B_4} &= B_{jk}d_id_j\delta_{kn} = \bigg\{[B]^T\{d\}\bigg\}\bigg\{d\bigg\}^T \label{C-S8-4}
\end{align}
\end{subequations}

\begin{subequations}
\label{C-S9}
\begin{align}
\mathrm{\bar C_1} &= C_{ijk}d_md_jd_k\delta_{mn} \nonumber \\
&= \left[\begin{array}{c}
\Bigg(\bigg\{[C_{1[JK]}]\{d\}\bigg\}^T\bigg\{d\bigg\}\Bigg)\Bigg\{d\Bigg\}^T \\
\Bigg(\bigg\{[C_{2[JK]}]\{d\}\bigg\}^T\bigg\{d\bigg\}\Bigg)\Bigg\{d\Bigg\}^T \\
\Bigg(\bigg\{[C_{3[JK]}]\{d\}\bigg\}^T\bigg\{d\bigg\}\Bigg)\Bigg\{d\Bigg\}^T
\end{array}\right]
\end{align}
\begin{align}
\mathrm{\bar C_2} = C_{ijk}d_j\delta_{kn} = \left[\begin{array}{c}
\bigg\{[C_{1[JN]}]^T\{d\}\bigg\}^T \\
\bigg\{[C_{2[JN]}]^T\{d\}\bigg\}^T \\
\bigg\{[C_{3[JN]}]^T\{d\}\bigg\}^T
\end{array}\right]
\end{align}
\begin{align}
\mathrm{\bar C_3} &= C_{kmj}d_md_id_j\delta_{kn} = \nonumber \\
&\left[\begin{array}{c}
\Bigg(\bigg\{[C_{1[MJ]}]\{d\}\bigg\}^T\bigg\{d\bigg\}\Bigg)\Bigg\{d\Bigg\} \\
\Bigg(\bigg\{[C_{2[MJ]}]\{d\}\bigg\}^T\bigg\{d\bigg\}\Bigg)\Bigg\{d\Bigg\} \\
\Bigg(\bigg\{[C_{3[MJ]}]\{d\}\bigg\}^T\bigg\{d\bigg\}\Bigg)\Bigg\{d\Bigg\}
\end{array}\right]^T
\end{align}
\begin{align}
\mathrm{\bar C_4} &= C_{kmi}d_m\delta_{kn} \nonumber \\
&= \left[\begin{array}{c}
\bigg\{[C_{1[MI]}]^T\{d\}\bigg\} \\
\bigg\{[C_{2[MI]}]^T\{d\}\bigg\} \\
\bigg\{[C_{3[MI]}]^T\{d\}\bigg\}
\end{array}\right]^T
\end{align}
\end{subequations}
where
\begin{subequations}
\label{C-S10}
\begin{align}
C_{1[JK]} = \left[\begin{array}{ccc}
C_{111} & C_{112} & C_{113} \\
C_{121} & C_{122} & C_{123} \\
C_{131} & C_{132} & C_{133} 
\end{array}\right]
\end{align}
\begin{align}
C_{2[JK]} = \left[\begin{array}{ccc}
C_{211} & C_{212} & C_{213} \\
C_{221} & C_{222} & C_{223} \\
C_{231} & C_{232} & C_{233} 
\end{array}\right]
\end{align}
\begin{align}
C_{3[JK]} = \left[\begin{array}{ccc}
C_{311} & C_{312} & C_{313} \\
C_{321} & C_{322} & C_{323} \\
C_{331} & C_{332} & C_{333} 
\end{array}\right]
\end{align}
\end{subequations}
Now, using Eq.~\eqref{C-S6},~\eqref{C-S7},~\eqref{C-S8},~\eqref{C-S9}, and~\eqref{C-S10}, Eq.~\eqref{C-S5} can be written as
\begin{align}\label{C-S11}
&\left[-[\mathrm{A}]k^2 + \frac{k^4}{16}[\mathrm{\bar B_1}] - \frac{k^4}{16}[\mathrm{\bar B_2}] - \frac{k^4}{16}[\mathrm{\bar B_3}] + \frac{k^4}{16}[\mathrm{\bar B_4}] \right.\nonumber \\
&\hspace{0.3in}\left.+ i\frac{k^3}{4}\bigg\{[\mathrm{\bar C_1}] - [\mathrm{\bar C_2}] - [\mathrm{\bar C_3}] + [\mathrm{\bar C_4}]\bigg\}\right]u_n \nonumber \\
&\hspace{0.3in}+ \rho\omega^2\delta_{in}u_n = 0
\end{align}
 
The wave propagation speed can be calculated from Eq.~\eqref{C-S11} in any direction by changing the propagation unit vector $\textbf{d}$. Because of the presence of third and fourth powers of $k$, some of the waves will be 
dispersive. Also, the matrices $[\bar C_1], [\bar C_2], [\bar C_3]$, and $[\bar C_4]$ have values such that waves are always 
dispersive, and there is no wave going out of bounds. Now, for general anisotropic media, we have the following restrictions on constitutive coefficients \citep{Ali2011, Ali2014thermo}:
\begin{align}
A_{ijkl} = A_{jikl} &= A_{klij} \nonumber \\
C_{ijk} &= C_{jik} \nonumber \\
B_{ij} &= B_{ji}\label{C-S12}
\end{align}
Equation \eqref{C-S12} leads to the reduction of 81, 27, and 9 coefficients to 21, 18, and 6 independent coefficients, respectively, such that
\begin{subequations}
\label{C-S13}
\begin{align}
A_{1111} &= A_{11} \nonumber \\
A_{2222} &= A_{22} \nonumber \\
A_{3333} &= A_{33}
\end{align}
\begin{align}
A_{3322} = A_{2233} &= A_{23} \nonumber \\
A_{3311} = A_{1133} &= A_{13} \nonumber \\
A_{1122} = A_{2211} &= A_{12}
\end{align}
\begin{align}
A_{2111} = A_{1211} = A_{1112} = A_{1121} &= A_{16} \nonumber \\
A_{3111} = A_{1311} = A_{1131} = A_{1113} &= A_{15} \nonumber \\
A_{3233} = A_{2333} = A_{3332} = A_{3323} &= A_{34} \nonumber \\
A_{2122} = A_{1222} = A_{2221} = A_{2212} &= A_{26} \nonumber \\
A_{3133} = A_{1333} = A_{3331} = A_{3313} &= A_{35} \nonumber \\
A_{3222} = A_{2322} = A_{2232} = A_{2223} &= A_{24}
\end{align}
\vspace{0pt}
\begin{align}
A_{3211} = A_{2311} = A_{1132} = A_{1123} &= A_{14} \nonumber \\
A_{3122} = A_{1322} = A_{2231} = A_{2213} &= A_{25} \nonumber \\
A_{2133} = A_{1233} = A_{3321} = A_{3312} &= A_{36} \nonumber \\
A_{3223} = A_{2323} = A_{2332} = A_{3232} &= A_{44} \nonumber \\
A_{3113} = A_{1313} = A_{1331} = A_{3131} &= A_{55} \nonumber \\
A_{2112} = A_{1212} = A_{1221} = A_{2121} &= A_{66}
\end{align}
\begin{align}
A_{2113} = A_{1213} = A_{1312} = A_{1321} &= A_{56} \nonumber \\
A_{3112} = A_{3121} = A_{1231} = A_{2131} &= A_{56} \nonumber \\
A_{2123} = A_{1223} = A_{2321} = A_{2312} &= A_{46} \nonumber \\
A_{3212} = A_{1232} = A_{3221} = A_{2132} &= A_{46} \nonumber \\
A_{3123} = A_{1323} = A_{2331} = A_{2313} &= A_{45} \nonumber \\
A_{3231} = A_{3213} = A_{3132} = A_{1332} &= A_{45}
\end{align}
\end{subequations}
\begin{subequations}
\label{C-S14}
\begin{align}
C_{211} &= C_{121} \nonumber \\
C_{212} &= C_{122} \nonumber \\
C_{213} &= C_{123}
\end{align}
\begin{align}
C_{311} &= C_{131} \nonumber \\
C_{312} &= C_{132} \nonumber \\
C_{313} &= C_{133}
\end{align}
\begin{align}
C_{321} &= C_{231} \nonumber \\
C_{322} &= C_{232} \nonumber \\
C_{323} &= C_{233} 
\end{align}
\end{subequations}
\begin{align}\label{C-S15}
B_{21} &= B_{12} \nonumber \\
B_{31} &= B_{13} \nonumber \\
B_{32} &= B_{23}
\end{align}
Thus, the most general anisotropic material requires 45 independent elastic coefficients.

Substituting Eq.~\eqref{C-S12},~\eqref{C-S13},~\eqref{C-S14}, and~\eqref{C-S15} in Eq.~\eqref{C-S11}, and solving for \textbf{d} = \{1 0 0\}, we obtain
\vspace{-0pt}
\begin{align}\label{C-S16}
\begin{small}\left[\begin{array}{ccc}\vspace{10pt}
-A_{11}k^2 + \rho\omega^2& -\frac{B_{12}k^4}{16} -\frac{iC_{112}k^3}{4} - A_{16}k^2 & -\frac{B_{13}k^4}{16} - \frac{i(C_{113} - C_{121} + C_{131})k^3}{4} - A_{15}k^2  \\ \vspace{10pt}
\frac{B_{12}k^4}{16} + \frac{iC_{112}k^3}{4} - A_{16}k^2 &  -\frac{B_{22}k^4}{16} - A_{66}k^2  + \rho\omega^2 & -\frac{B_{23}k^4}{16} + \frac{i(C_{122} - C_{123})k^3}{4} - A_{56}k^2 \\
\frac{B_{13}k^4}{16} + \frac{iC_{113}k^3}{4} - A_{15}k^2 & -\frac{B_{23}k^4}{16} + \frac{i(C_{123} - C_{132})k^3}{4} - A_{56}k^2 & -\frac{B_{33}k^4}{16} + \frac{i(C_{123} - C_{133})k^3}{4} - A_{55}k^2  + \rho\omega^2
\end{array}\right]\end{small}u_n = 0 \nonumber \\
\end{align}

\vspace{0pt}
Now, by solving Eq.~\eqref{C-S16}, we can determine the wave velocity. Similarly, for other directions, Eq.~\eqref{C-S11} can be evaluated. In the next section, we will specifically focus on cubic materials, and illustrate wave propagation in all three directions of primary interest.

\section{Wave Propagation in Cubic Crystals}
For cubic crystals, we have additional symmetry directions, and the number of total independent constants in consistent couple stress theory \citep{Ali2011} reduces dramatically from the 45 present in the most general anisotropic case.  The fourth order $A$ tensor reduces to its form in classical elasticity with three independent material constants for cubic crystals, the second order $B$ tensor becomes isotropic having just a single material coefficient, while the third order $C$ tensor vanishes for all centrosymmetric crystals.  Thus, the total number of independent constants for cubic single crystals is four, which are defined as follows:
\begin{subequations}
\label{C-S17}
\begin{align}
A_{1111} &= A_{11} \nonumber \\
A_{2222} &= A_{11} \nonumber \\
A_{3333} &= A_{11}
\end{align}
\begin{align}
A_{3322} = A_{2233} &= A_{12} \nonumber \\
A_{3311} = A_{1133} &= A_{12} \nonumber \\
A_{1122} = A_{2211} &= A_{12}
\end{align}
\begin{align}
A_{2111} = A_{1211} = A_{1112} = A_{1121} &= A_{16} = 0 \nonumber \\
A_{3111} = A_{1311} = A_{1131} = A_{1113} &= A_{15} = 0  \nonumber \\
A_{3233} = A_{2333} = A_{3332} = A_{3323} &= A_{34} = 0  \nonumber \\
A_{2122} = A_{1222} = A_{2221} = A_{2212} &= A_{26} = 0  \nonumber \\
A_{3133} = A_{1333} = A_{3331} = A_{3313} &= A_{35} = 0  \nonumber \\
A_{3222} = A_{2322} = A_{2232} = A_{2223} &= A_{24} = 0 
\end{align}
\begin{align}
A_{3211} = A_{2311} = A_{1132} = A_{1123} &= A_{14}  = 0 \nonumber \\
A_{3122} = A_{1322} = A_{2231} = A_{2213} &= A_{25}  = 0 \nonumber \\
A_{2133} = A_{1233} = A_{3321} = A_{3312} &= A_{36}  = 0 \nonumber \\
A_{3223} = A_{2323} = A_{2332} = A_{3232} &= A_{44} \nonumber \\
A_{3113} = A_{1313} = A_{1331} = A_{3131} &= A_{44} \nonumber \\
A_{2112} = A_{1212} = A_{1221} = A_{2121} &= A_{44}
\end{align}
\begin{align}
A_{2113} = A_{1213} = A_{1312} = A_{1321} &= A_{56} = 0  \nonumber \\
 A_{3112} = A_{3121} = A_{1231} = A_{2131} &= A_{56} = 0  \nonumber \\
A_{2123} = A_{1223} = A_{2321} = A_{2312} &= A_{46} = 0  \nonumber \\ 
A_{3212} = A_{1232} = A_{3221} = A_{2132} &= A_{46} = 0  \nonumber \\
A_{3123} = A_{1323} = A_{2331} = A_{2313} &= A_{45} = 0 \nonumber \\
 A_{3231} = A_{3213} = A_{3132} = A_{1332} &= A_{45} = 0 
\end{align}
\end{subequations}
\vspace{-10pt}
\begin{subequations}
\label{C-S18}
\begin{align}
C_{211} &= C_{121} = 0  \nonumber \\
C_{212} &= C_{122} = 0  \nonumber \\
C_{213} &= C_{123} = 0
\end{align}
\begin{align}
C_{311} &= C_{131} = 0  \nonumber \\
C_{312} &= C_{132} = 0  \nonumber \\
C_{313} &= C_{133} = 0
\end{align}
\begin{align}
C_{321} &= C_{231} = 0  \nonumber \\
C_{322} &= C_{232} = 0  \nonumber \\
C_{323} &= C_{233} = 0  
\end{align}
\begin{align}
C_{111} &= C_{112} = C_{113} = 0  \nonumber \\
C_{221} &= C_{222} = C_{223} = 0  \nonumber \\
C_{331} &= C_{332} = C_{333} = 0  
\end{align}
\end{subequations}
\begin{align}\label{C-S19}
B_{11} = B_{22} = B_{33} &= 16\eta \nonumber \\
B_{21} = B_{12} &= 0  \nonumber \\
B_{31} = B_{13} &= 0  \nonumber \\
B_{32} = B_{23} &= 0 
\end{align}
Thus, $A_{11}$, $A_{12}$, $A_{44}$ and the size-dependent couple stress material parameter $\eta$ must be determined.

Substituting Eq.~\eqref{C-S17},~\eqref{C-S18}, and~\eqref{C-S19} in Eq.~\eqref{C-S11}, and solving for \textbf{d} = \{1 0 0\}, we find
\begin{align}\label{C-S20}
\left[\begin{array}{ccc}\vspace{10pt}
-A_{11}k^2 + \rho\omega^2 & 0 & 0  \\ \vspace{10pt}
0 &  -\eta k^4 - A_{44}k^2  + \rho\omega^2 & 0 \\
 0 & 0 & -\eta k^4 - A_{44}k^2  + \rho\omega^2
\end{array}\right]u_n = 0
\end{align}
Now, solving Eq.~\eqref{C-S20}, the wave velocities in the longitudinal and transverse directions are given by
\begin{align}\label{C-S21}
\rho\omega_{\{100\}L}^2 &= A_{11}k^2 \nonumber \\
\rho\omega_{\{100\}T1}^2 &= \eta k^4 + A_{44}k^2 \nonumber \\
\rho\omega_{\{100\}T2}^2 &= \eta k^4 + A_{44}k^2
\end{align}
Similarly, substituting Eq.~\eqref{C-S17},~\eqref{C-S18}, and~\eqref{C-S19} in Eq.~\eqref{C-S11}, and solving for \textbf{d} = $\frac{1}{\sqrt{2}}$\{1 1 0\}, we obtain
\begin{align}\label{C-S22}
&\begin{normalsize}\left[\begin{array}{ccc}\vspace{10pt}
-\frac{\eta k^4}{2} - \frac{(A_{11}+A_{44})}{2}k^2 + \rho\omega^2 & \frac{\eta k^4}{2} - \frac{(A_{12}+A_{44})}{2}k^2 & 0  \\ \vspace{10pt}
\frac{\eta k^4}{2} - \frac{(A_{12}+A_{44})}{2}k^2 &  -\frac{\eta k^4}{2} - \frac{(A_{11}+A_{44})}{2}k^2 + \rho\omega^2 & 0 \\
 0 & 0 & -\eta k^4 - A_{44}k^2  + \rho\omega^2
\end{array}\right]u_n = 0\end{normalsize}
\end{align}
Next, solving Eq.~\eqref{C-S22}, the wave velocities in the longitudinal and transverse directions are given by
\begin{align}\label{C-S23}
\rho\omega_{\{110\}L}^2 &= \frac{(A_{11}+A_{12})}{2}k^2 + A_{44}k^2 \nonumber \\
\rho\omega_{\{110\}T1}^2 &= \eta k^4 + \frac{(A_{11}-A_{12})}{2}k^2 \nonumber \\
\rho\omega_{\{110\}T2}^2 &= \eta k^4 + A_{44}k^2
\end{align}
Similarly, substituting Eq.~\eqref{C-S17},~\eqref{C-S18}, and~\eqref{C-S19} in Eq.~\eqref{C-S11}, and solving for \textbf{d} = $\frac{1}{\sqrt{3}}$\{1 1 1\}, we may write
\begin{align}\label{C-S24}
\begin{normalsize}\left[\begin{array}{ccc}\vspace{10pt}
-\frac{2\eta k^4}{3} - \frac{(A_{11}+2 A_{44})}{3}k^2 + \rho\omega^2 & \frac{\eta k^4}{3} - \frac{(A_{12}+A_{44})}{3}k^2 & \frac{\eta k^4}{3} - \frac{(A_{12}+A_{44})}{3}k^2  \\ \vspace{10pt}
\frac{\eta k^4}{3} - \frac{(A_{12}+A_{44})}{3}k^2 &  -\frac{2\eta k^4}{3} - \frac{(A_{11}+2 A_{44})}{3}k^2 + \rho\omega^2 & \frac{\eta k^4}{3} - \frac{(A_{12}+A_{44})}{3}k^2 \\
 \frac{\eta k^4}{3} - \frac{(A_{12}+A_{44})}{3}k^2 & \frac{\eta k^4}{3} - \frac{(A_{12}+A_{44})}{3}k^2 & -\frac{2\eta k^4}{3} - \frac{(A_{11}+2 A_{44})}{3}k^2 + \rho\omega^2
\end{array}\right]\end{normalsize}u_n = 0
\end{align}
\vspace{12pt}
Finally, solving Eq.~\eqref{C-S24}, the wave velocities in the longitudinal and transverse direction are given by
\begin{align}\label{C-S25}
\rho\omega_{\{111\}L}^2 &= \frac{(A_{11}+2A_{12} + 4A_{44})}{3}k^2 \nonumber \\
\rho\omega_{\{111\}T1}^2 &= \eta k^4 + \frac{(A_{11}+A_{44}-A_{12})}{3}k^2 \nonumber \\
\rho\omega_{\{111\}T2}^2 &= \eta k^4 + \frac{(A_{11}+A_{44}-A_{12})}{3}k^2
\end{align}

Before turning to the evaluation of the elastic constants for cubic crystals, let us briefly revisit the case of isotropic materials.  For this case, the elastic constants satisfy the relation $A_{44}=( A_{11} - A_{12} )/2$.  Substituting this into Eqs.~\eqref{C-S21}, \eqref{C-S23} and \eqref{C-S25}, we find that the wave velocities are given by
\begin{align}\label{C-S21iso}
\rho\omega_{L}^2 &= A_{11}k^2 \nonumber \\
\rho\omega_{T1}^2 &= \eta k^4 + A_{44}k^2 \nonumber \\
\rho\omega_{T2}^2 &= \eta k^4 + A_{44}k^2
\end{align}
independent of the direction \textbf{d}, as expected for isotropy.  The longitudinal wave is non-dispersive, exactly as in the classical case.  On the other hand, from \eqref{C-S21iso}, the two transverse waves are dispersive in consistent couple stress theory due to the contributions from curvature.

\vspace{0pt}
\section{Evaluation of the Four Adiabatic Elastic Constants for Cubic Crystals}

From the literature, we have collected the wave velocity data for face-centered cubic (Cu), body-centered cubic (CuZn), and simple cubic (NaCl and KCl) from ultrasonic measurements with a frequency of 12MHz in different directions of the single crystal \citep{Lazarus1949}. The adiabatic bulk moduli of Cu, NaCl, and KCl have been obtained from data of \cite{Raman-Krishna9-1955}, while that of CuZn has been obtained from \cite{Good1941}. The following wave velocity measurements are available from the experiments:\\

\indent${V_1}$ = velocity of longitudinal wave in (100) crystal in \{100\} direction,\\ 
\indent${V_2}$ = velocity of longitudinal wave in (100) crystal in \{110\} direction,\\ 
\indent${V_3}$ = velocity of transverse wave in (100) crystal in \{100\} direction,\\ 
\indent${V_4}$ = velocity of transverse wave in (110) crystal in \{1$\overline{1}$0\} direction, and\\
\indent${V_5}$ = velocity of transverse wave in (110) crystal in \{100\} direction.\\

Now for the calculation of elastic constants, we have a system of $n$ equations and $K$ variables, where $n>K$. There are multiple ways to solve the problem. The most frequently used ones are numerical optimization techniques and least squares estimation. Numerical optimization techniques provide excellent ways to solve the deterministic problem. However, there is one difficulty with the formulation. There are experimental errors in the data of wave velocity and adiabatic bulk modulus. The experimental error can be given as an input to the optimization problem in the form of constraints; however that will force us to use the uniform distribution of the value of wave velocity and bulk modulus over the error range. The least squares approach does not take errors into the formulation. With a weighted least squares approach, we can assign weights to experimental measurements, but error weights are not available in the present case. Another scientific approach to obtain the solution is to make use of a minimum variance (MV) or maximum likelihood (ML) approach \citep{Crassidis2012} to estimate the value of the elastic constants. If we formulate the problem in a linear system of equations with Gaussian distribution of error, both MV and ML approaches produce the same results. Even with the weighted least squares method, the results are the same as MV and ML, if weights are assigned equal to the value of the inverse of the variance of the experimental measurements. 

In the present work, the observed experimental data is assigned a Gaussian distribution with mean as the observed experimental value and standard deviation ($\sigma$) as the square root of the variance of the error. Thus, MV, ML and weighted least squares, all produce the same results. Using Eq.~\eqref{C-S21} and~\eqref{C-S23}, the system of linear equations can be expressed as follows:
\begin{align}\label{C-S26}
\bar{\textbf{y}} = H\mathrm{\textbf{x}} + {\nu}
\end{align}
where

\begin{subequations}
\label{C-S27}
\begin{align}
\bar{\textbf{y}} = \left[\begin{array}{c}\vspace{10pt}
\rho V_1^2 \\ \vspace{10pt}2\rho V_2^2 \\ \vspace{10pt}\rho\left(\frac{V_3+V_5}{2}\right)^2 \\ \vspace{10pt}2\rho V_4^2 \\ K_{ad}
\end{array}\right]
\end{align} 
\begin{align}
H = \left[\begin{array}{cccc}\vspace{10pt}
1 & 0 & 0 & 0\\ \vspace{10pt}
1 & 1 & 2 & 0\\ \vspace{10pt}
0 & 0 & 1 & k_{v3}^2\\ \vspace{10pt}
1 & -1 & 0 & 2k_{v4}^2 \\
1/3 & 2/3 & 0 & 0
\end{array}\right]; \text{and } \mathrm{\textbf{x}} = \left[\begin{array}{c}\vspace{10pt}
A_{11} \\ \vspace{10pt}A_{12} \\ \vspace{10pt}A_{44} \\ \eta
\end{array}\right]
\end{align}
\end{subequations}
and ${\nu}$ is the error in experimental observations of $\bar{\textbf{y}}$, $K_{ad}$ is the adiabatic bulk modulus of the material and $k_{v3}$ and $k_{v4}$ are the wave numbers of the waves traveling with velocity $V_3$ and $V_4$, respectively. The MV or ML solution of the linear system of equation is given by
\begin{align}\label{C-S28}
\hat{\mathrm{\textbf{x}}} = (H^{T}R^{-1}H)^{-1}H^TR^{-1}\bar{\textbf{y}}
\end{align}
and the error covariance matrix $P$ of the optimal solution of elastic constants $\hat{\mathrm{\textbf{x}}}$ is given by
\begin{align}\label{C-S29}
P = (H^{T}R^{-1}H)^{-1}
\end{align}
where $R = E[\nu\nu^T]$ is the observation error covariance matrix given by 
\begin{align}\label{C-S30}
R = \left[\begin{array}{ccccc}
\nu_1^2 & 0 & 0 & 0 & 0\\
0 & \nu_2^2 & 0 & 0 & 0\\
0 & 0 & \nu_3^2 & 0 & 0\\
0 & 0 & 0 & \nu_4^2 & 0\\
0 & 0 & 0& 0 & \nu_5^2
\end{array}\right]
\end{align}
where measurements have been assumed to be independent of each other.

\cite{Lazarus1949} has stated the error parameter $\sigma$ in the calculation of elastic constants to be 0.2\% of the measured value. There is no mention of error in measurements of the adiabatic bulk modulus by \cite{Raman-Krishna9-1955} and by \cite{Good1941}; however they mention the measurements to be very accurate and reliable, thus $\sigma$ in the adiabatic bulk modulus has been assumed to be 0.1\% of the value of the experimental data. (The sensitivity of the assumed error is described later in the section.) Numerical values of error variances are listed below for individual cases. Unless otherwise specified, the units of $A_{11}$, $A_{12}$, and $A_{44}$ are $N/m^2$, while $\eta$ is expressed in $N$.
\subsection{Simple Cubic Crystal -- NaCl}
The experimental measurements of wave velocity and observation error covariance matrix for NaCl are as follows \citep{Lazarus1949, Raman-Krishna9-1955}:
\begin{align}
\begin{array}{lcr}
V_1 & = & \text{4766 }m/sec  \\
V_2 & = & \text{4513 }m/sec  \\
V_3 & = & \text{2434 }m/sec  \\
V_4 & = & \text{2920 }m/sec  \\
V_5 & = & \text{2440 }m/sec  \\
K_{ad} & = & 2.52\times10^{10}\text{ }N/m^2  \\
\rho & = & \text{2162 $Kg/m^3$} 
\end{array}
\end{align}
\begin{small}
\begin{align}\label{C-S31}
R = 10^{16}\left[\begin{array}{ccccc}
0.9643     &    0    &     0     &    0     &    0 \\
         0   & 3.0311     &    0     &    0     &    0 \\
         0    &     0   & 0.0659    &     0     &    0 \\
         0    &     0    &     0  &  1.5055   &      0 \\
         0    &     0    &     0    &     0  &  0.0635
\end{array}\right]
\end{align}
\end{small}

\noindent Using Eq.~\eqref{C-S27},~\eqref{C-S28}, and~\eqref{C-S29}, the optimal values of the elastic constants and error covariance matrix are calculated as
\begin{align}\label{C-S32}
\hat{\mathrm{\textbf{x}}} = \left[\begin{array}{c}
A_{11}\\A_{12}\\A_{44}\\\eta
\end{array}\right] = \left[\begin{array}{c}
4.94\times 10^{10}\\1.31\times 10^{10}\\1.26\times 10^{10}\\ 9.68
\end{array}\right]
\end{align}
  
 \begin{small}
 \begin{align}\label{C-S33}
&&P = 10^{15}\left[\begin{array}{cccc}
4.6848 &   -2.2210 &    1.6990 &    -0.8039\times 10^{-7} \\
    &    2.2457 &   -1.6645 &     0.7398\times 10^{-7} \\
     &    &    5.1312 &  -2.0240\times 10^{-7} \\
  \multicolumn{2}{c}{\text{\smash{\raisebox{1.5ex}{Sym.}}}}& &     9.0634\times 10^{-15}
\end{array}\right]
\end{align}
 \end{small}
 
It should be noted that the error covariance matrix is fully populated. This is because $A_{11}$, $A_{12}$, $A_{44}$, and $\eta$ all depend on each other for given experimental data as defined by Eq.~\eqref{C-S26} and~\eqref{C-S27}. Thus, error in one of the elastic constants affects the error in any other. Also, note the importance of using an MV or ML approach.  Along with solving for the elastic constants, we are able to calculate the error covariance between any two elastic constants, which is not possible with constrained optimization techniques. The mean $\pm$ standard deviation of the adiabatic elastic constants is calculated as
\begin{align}\label{C-S34}
{\mathrm{\textbf{x}}} = \left[\begin{array}{c}
A_{11} \pm \sigma \\A_{12}\pm \sigma\\A_{44}\pm \sigma\\\eta\pm \sigma
\end{array}\right] = \left[\begin{array}{c}
4.94\times 10^{10}\pm 6.85\times 10^7\\1.31\times 10^{10}\pm 4.98\times 10^7\\1.26\times 10^{10}\pm 7.16\times 10^7\\ 9.68\pm 3.01
\end{array}\right]
\end{align}
\subsection{Simple Cubic Crystal -- KCl}
The experimental measurements of wave velocity and observation error covariance matrix for KCl are as follows \citep{Lazarus1949, Raman-Krishna9-1955}:
\begin{align}
\begin{array}{lcr}
V_1 = \text{4541 }m/sec \ \\
V_2 = \text{3896 }m/sec  \\
V_3 = \text{1781 }m/sec  \\
V_4 = \text{2921 }m/sec  \\
V_5 = \text{1784 }m/sec  \\
K_{ad} = 1.87\times10^{10}\text{ }N/m^2 \\
\rho = \text{1986 $Kg/m^3$}  
\end{array}
\end{align}
\begin{small}
\begin{align}\label{C-S35}
R = 10^{16}\left[\begin{array}{ccccc}
0.6708     &    0    &     0     &    0     &    0 \\
         0   & 1.4701     &    0     &    0     &    0 \\
         0    &     0   & 0.0159    &     0     &    0 \\
         0    &     0    &     0  &  0.9217   &      0 \\
         0    &     0    &     0    &     0  &  0.0348
\end{array}\right]
\end{align}
\end{small}

\noindent Using Eq.~\eqref{C-S27},~\eqref{C-S28}, and~\eqref{C-S29}, the optimal values of the elastic constants and error covariance matrix are calculated as
\begin{align}\label{C-S36}
\hat{\mathrm{\textbf{x}}} = \left[\begin{array}{c}
A_{11}\\A_{12}\\A_{44}\\\eta
\end{array}\right] = \left[\begin{array}{c}
4.10\times 10^{10}\\0.75\times 10^{10}\\0.59\times 10^{10}\\ 9.17
\end{array}\right]
\end{align}
\begin{small}
\begin{align}\label{C-S37}
&&P = 10^{15}\left[\begin{array}{cccc}
2.5437 &   -1.0908 &    0.2798 &    -0.7124\times 10^{-8} \\
   &    1.2296 &   -0.6163 &     1.4097\times 10^{-8} \\
   &  &   3.3706 &  -7.3794\times 10^{-8} \\
    \multicolumn{2}{c}{\text{\smash{\raisebox{1.5ex}{Sym.}}}}     &   &    1.6926\times 10^{-15}
\end{array}\right]
\end{align}
\end{small}
The mean $\pm$ standard deviation of the adiabatic elastic constants is calculated as
\begin{align}\label{C-S38}
{\mathrm{\textbf{x}}} = \left[\begin{array}{c}
A_{11} \pm \sigma \\A_{12}\pm \sigma\\A_{44}\pm \sigma\\\eta\pm \sigma
\end{array}\right] = \left[\begin{array}{c}
4.10\times 10^{10}\pm 5.04\times 10^7\\0.75\times 10^{10}\pm 3.51\times 10^7\\0.59\times 10^{10}\pm 5.81\times 10^7\\ 9.17\pm 1.30
\end{array}\right]
\end{align}

\subsection{Face-Centered Cubic Crystal -- Cu}
The experimental measurements of wave velocity and observation error covariance matrix for Cu are as follows \citep{Lazarus1949, Raman-Krishna9-1955}:
\begin{align}
\begin{array}{lcr}
V_1 & = & \text{4373 }m/sec  \\
V_2 & = & \text{4982 }m/sec  \\
V_3 & = & \text{2905 }m/sec  \\
V_4 & = & \text{1621 }m/sec  \\
V_5 & = & \text{2913 }m/sec \\
K_{ad} & = & 14.18\times10^{10}\text{ }N/m^2  \\
\rho & = & \text{8941 $Kg/m^3$} 
\end{array}
\end{align}
\begin{small}
\begin{align}\label{C-S39}
R = 10^{17}\left[\begin{array}{ccccc}
1.1694     &    0    &     0     &    0     &    0 \\
         0   & 7.9639     &    0     &    0     &    0 \\
         0    &     0   & 0.2290    &     0     &    0 \\
         0    &     0    &     0  &  3.4782   &      0 \\
         0    &     0    &     0    &     0  &  0.2011
\end{array}\right]
\end{align}
\end{small}

\noindent Using Eq.~\eqref{C-S27},~\eqref{C-S28}, and~\eqref{C-S29}, the optimal values of the elastic constants and error covariance matrix are calculated as
\begin{align}\label{C-S40}
\hat{\mathrm{\textbf{x}}} = \left[\begin{array}{c}
A_{11}\\A_{12}\\A_{44}\\\eta
\end{array}\right] = \left[\begin{array}{c}
1.71\times 10^{11}\\1.27\times 10^{11}\\0.75\times 10^{11}\\ 39.35
\end{array}\right]
\end{align}
\begin{small}
\begin{align}\label{C-S41}
&&P = 10^{17}\left[\begin{array}{cccc}
1.0479  &   -0.5574 &    0.1826 &    -1.3584\times 10^{-8} \\
   &    0.7386 &   -0.1867 &     1.1586\times 10^{-8} \\
     &   &   0.3206 &  -0.7519\times 10^{-8} \\
    \multicolumn{2}{c}{\text{\smash{\raisebox{1.5ex}{Sym.}}}}     &   &   49.8049\times 10^{-17}
\end{array}\right]
\end{align}
\end{small}
The mean $\pm$ standard deviation of the adiabatic elastic constants is calculated as
\begin{align}\label{C-S42}
{\mathrm{\textbf{x}}} = \left[\begin{array}{c}
A_{11} \pm \sigma \\A_{12}\pm \sigma\\A_{44}\pm \sigma\\\eta\pm \sigma
\end{array}\right] = \left[\begin{array}{c}
1.71\times 10^{11}\pm 3.24\times 10^8\\1.27\times 10^{11}\pm 2.72\times 10^8\\0.75\times 10^{11}\pm 1.79\times 10^8\\ 39.35\pm 7.06
\end{array}\right]
\end{align}

\subsection{Body-Centered Cubic Crystal -- CuZn}
The experimental measurements of wave velocity and observation error covariance matrix for CuZn are as follows \citep{Lazarus1949, Good1941}:
\begin{align}
\begin{array}{lcr}
V_1 & = & \text{3942 }m/sec  \\
V_2 & = & \text{4931 }m/sec  \\
V_3 & = & \text{3151 }m/sec  \\
V_4 & = & \text{1083 }m/sec  \\
V_5 & = & \text{3152 }m/sec \\
K_{ad} & = & 11.93\times10^{10}\text{ }N/m^2  \\
\rho & = & \text{8304 $Kg/m^3$} 
\end{array}
\end{align}
\begin{small}
\begin{align}\label{C-S43}
R = 10^{17}\left[\begin{array}{ccccc}
0.6660     &    0    &     0     &    0     &    0 \\
         0   & 6.5186     &    0     &    0     &    0 \\
         0    &     0   & 0.2721    &     0     &    0 \\
         0    &     0    &     0  &  2.2799   &      0 \\
         0    &     0    &     0    &     0  &  0.1422
\end{array}\right]
\end{align}
\end{small}

\noindent Using Eq.~\eqref{C-S27},~\eqref{C-S28}, and~\eqref{C-S29}, the optimal values of the elastic constants and error covariance matrix are calculated as
\begin{align}\label{C-S44}
\hat{\mathrm{\textbf{x}}} = \left[\begin{array}{c}
A_{11}\\A_{12}\\A_{44}\\\eta
\end{array}\right] = \left[\begin{array}{c}
1.29\times 10^{11}\\1.14\times 10^{11}\\0.82\times 10^{11}\\ 20.90
\end{array}\right]
\end{align}
\begin{small}
\begin{align}\label{C-S45}
&&P = 10^{16}\left[\begin{array}{cccc}
6.4121 &   -3.3614 &   0.2689 &    -3.9199\times 10^{-8} \\
   &    4.8613  &   -0.5244 &     3.3563\times 10^{-8} \\
     &   &   2.4485 &  -0.7934\times 10^{-8} \\
  \multicolumn{2}{c}{\text{\smash{\raisebox{1.5ex}{Sym.}}}}     &   &    6.7083\times 10^{-16}
\end{array}\right]
\end{align}
\end{small}
The mean $\pm$ standard deviation of the adiabatic elastic constants is calculated as
\begin{align}\label{C-S46}
{\mathrm{\textbf{x}}} = \left[\begin{array}{c}
A_{11} \pm \sigma \\A_{12}\pm \sigma\\A_{44}\pm \sigma\\\eta\pm \sigma
\end{array}\right] = \left[\begin{array}{c}
1.29\times 10^{11}\pm 2.53\times 10^8\\1.14\times 10^{11}\pm 2.21\times 10^8\\0.82\times 10^{11}\pm 1.57\times 10^8\\ 20.90\pm 2.59
\end{array}\right]
\end{align}

\subsection{Summary of Elastic Constant Estimations}
With the present stochastic analysis, we have been able to estimate the value of the elastic constants within an error range and also establish the error covariance matrix that can indicate the accuracy of one elastic constant that has a high dependence on the value of another elastic constant.  From all the above calculated adiabatic elastic constants, it should be noted that small errors in experimental measurements of wave velocity and adiabatic bulk modulus lead to relatively large errors in results for the couple stress material parameter $\eta$.  More specifically, the coefficient of variation in $\eta$ ranges from approximately 12\% for CuZn to 31\% for NaCl.  

Although the reported adiabatic bulk modulus data is quite accurate, we checked the sensitivity associated with its variability by increasing the assumed error dramatically from 0.1\% to 0.5\%.  All elastic constants remained nearly constant, except for the value of $\eta$, which may change to 30\% of its value with 0.1\% error for NaCl and by lesser amounts for the other three materials.  Thus, for 0.5\% bulk modulus error in NaCl, $\eta$ lies within approximately $\pm2\sigma$ of its mean value at 0.1\% error.

If we wish to constrain the value of the couple stress elastic constant within tighter bounds, more precise measurements of wave velocity and adiabatic bulk modulus may be required.  Alternatively, we would need to measure the wave velocities at higher frequencies, such that the contributions of the size-dependent terms are of the same order of magnitude as the classical size-independent elastic terms in our material property estimation algorithm.

\section{Discussion}
\subsection{Relative Likelihood of Elasticity Theories}
So far, we have given theoretical arguments in favor of size-dependent consistent couple stress theory \citep{Ali2011} as the true representation of solid continua at high frequencies/small scales, which then requires four elastic constants to describe the linear elastic response of cubic crystals.  In this discussion, 
we also will consider classical continuum mechanics theory, which indeed produces waves that are physically present in cubic single crystals and have been experimentally verified.  The difference between classical continuum mechanics and couple stress theory arises at high frequencies/small scales, where there have been very few measurements.  We note here that by high frequencies we mean frequencies at which the assumption of a continuum is valid, although at present we still do not know the frequency at which this assumption starts to fail.  Most of the measurements of wave velocity to determine elastic constants at high frequencies are in the ultrasonic region (MHz).  In this section, based on the available data, we will investigate whether consistent couple stress theory is indeed better than Cauchy elasticity theory at high frequencies/small scale to represent the behavior of the four different cubic single crystals examined in the previous section.

%

Conventionally, it is possible to increase the likelihood of one model by increasing the number of parameters. However, this may result in overfitting. It is important also to penalize the complexity of the model along with rewarding the likelihood or approximation capability of the model for model selection. According to information theory, there are many ways to evaluate the performance of one model relative to another; the most prominent approaches are the Bayesian information criterion (BIC) and the Akaike information criterion (AIC) \citep{YYang2005}. The advantage of using the BIC and AIC lies in the fact that the model performance depends upon the information extracted from a given number of parameters, and it does not necessarily improve with an increase in the number of parameters. Thus, for example, a six parameter model may behave poorly in comparison to a two parameter model. Both BIC and AIC resolve the problem of overfitting by introducing a penalty term for the number of parameters in the model. We will make use of both of these criteria to investigate the superiority (in statistics language, very strong evidence against classical continuum mechanics or relative likelihood) of couple stress theory over classical continuum mechanics at high frequencies/small scale. 

The models here are couple stress theory  (Model A) and classical continuum mechanics theory (Model B). First we need to represent both models in a similar mathematical form.  As shown in the previous section, couple stress theory can be represented as
\begin{align}\label{C-S47}
\bar{\textbf{y}}_A =  H_A\mathrm{\textbf{x}_A} + {\nu_A}
\end{align}
where

\begin{subequations}
\label{C-S48}
\begin{align}
\bar{\textbf{y}}_A = \left[\begin{array}{c}\vspace{10pt}
\rho V_1^2 \\ \vspace{10pt}2\rho V_2^2 \\ \vspace{10pt}\rho\left(\frac{V_3+V_5}{2}\right)^2 \\ \vspace{10pt}2\rho V_4^2 \\ K_{ad}
\end{array}\right]
\end{align}

\begin{align}
H_A = \left[\begin{array}{cccc}\vspace{10pt}
1 & 0 & 0 & 0\\ \vspace{10pt}
1 & 1 & 2 & 0\\ \vspace{10pt}
0 & 0 & 1 & k_{v3}^2\\ \vspace{10pt}
1 & -1 & 0 & 2k_{v4}^2 \\
1/3 & 2/3 & 0 & 0
\end{array}\right]; \text{and } \mathrm{\textbf{x}_A} = \left[\begin{array}{c}\vspace{10pt}
A_{11} \\ \vspace{10pt}A_{12} \\ \vspace{10pt}A_{44} \\ \eta
\end{array}\right]
\end{align}
\end{subequations}
and ${\nu_A}$ is the error in the experimental observation of $\bar{\textbf{y}}_A$. 

Similarly, for classical elasticity, the mathematical model can be represented as
\begin{align}\label{C-S49}
\bar{\textbf{y}}_B =  H_B\mathrm{\textbf{x}_B} + {\nu_B}
\end{align}
where
\begin{subequations}
\label{C-S50}
\begin{align}
\bar{\textbf{y}}_B = \left[\begin{array}{c}\vspace{10pt}
\rho V_1^2 \\ \vspace{10pt}2\rho V_2^2 \\ \vspace{10pt}\rho\left(\frac{V_3+V_5}{2}\right)^2 \\ \vspace{10pt}2\rho V_4^2 \\ K_{ad}
\end{array}\right]
\end{align}
\begin{align}
H_B = \left[\begin{array}{ccc}\vspace{10pt}
1 & 0 & 0\\ \vspace{10pt}
1 & 1 & 2\\ \vspace{10pt}
0 & 0 & 1\\ \vspace{10pt}
1 & -1 & 0 \\
1/3 & 2/3 & 0
\end{array}\right]; \text{and } \mathrm{\textbf{x}_B} = \left[\begin{array}{c}\vspace{10pt}
A_{11} \\ \vspace{10pt}A_{12} \\ A_{44}
\end{array}\right]
\end{align}
\end{subequations}
and ${\nu_B}$ is the error in the experimental observation of $\bar{\textbf{y}}_B$. 


One thing to note here is that we are forcing both models to represent the adiabatic bulk modulus, as in their memoir \cite{Raman-Krishna9-1955} clearly pointed out the limitation of classical continuum mechanics theory to explain the adiabatic bulk modulus at high frequencies. Thus, following this approach, we are ensuring that the models capture both the wave velocities and the adiabatic bulk modulus. 

The solution for the couple stress theory model has been shown in the previous section.  However, while solving for the three elastic constants in the classical elasticity model (Eq.~\eqref{C-S49}, and~\eqref{C-S50}), we have observed that the model is unable to satisfy both wave velocities and adiabatic bulk modulus simultaneously.  In fact, the classical model produces large residual errors ($\bar{\textbf{y}}_B - H\hat{\mathrm{\textbf{x}}}_B$) that are as high as six times the standard deviation of the error in experimental measurements of $\bar{\textbf{y}}_B$, i.e., $\pm 6\sigma$, thus indicating the inconsistency in the theory at high frequencies/small length scales. Tables~\ref{T1} and~\ref{T2} show the elastic constants, and the maximum residual error $MRE = max(|\bar{\textbf{y}}_{i} - H\hat{\mathrm{\textbf{x}}}_{i}|/\sigma_{\bar{\textbf{y}}_{i}})$, calculated from classical elasticity and couple stress theory, where $\sigma_{\bar{\textbf{y}}_i}$ is the square root of the observation error covariance matrix $R$ defined in the last section. See Table~\ref{T} for residual error of each measurement data for both theories. From Table~\ref{T1}, we can observe that by forcing the elastic constants to satisfy the adiabatic bulk modulus, the values of $A_{11}$, $A_{12}$, and $A_{44}$ are the same within the first decimal with the couple stress theory results; however, now without the couple stress parameter $\eta$, these values do not satisfy the wave velocities within an error bound as indicated by maximum residual error. The values of elastic constants without forcing the adiabatic bulk modulus have been calculated previously \citep{Galt1948, Huntington1947, Raman-Krishna9-1955, Lazarus1949}; however as pointed out by \cite{Raman-Krishna9-1955}, these elastic constants clearly do not represent well the adiabatic bulk modulus.


\begin{table}
\begin{center}
  \caption{Elastic constants and maximum residual error for NaCl, KCl, Cu, and CuZn using classical elasticity theory.}
  \vspace{10pt}
  \label{T1}
  \begin{tabular}{ccccc}
  \hline
  \textbf{} & \textbf{NaCl} & \textbf{KCl} & \textbf{Cu} & \textbf{CuZn} \\
  \hline
   $A_{11}\ (Pa)$ & $4.95\times 10^{10}$   & $4.11\times 10^{10}$ & $1.72\times 10^{11}$ &  $1.30\times 10^{11}$\\
   $A_{12}\ (Pa)$ & $1.30\times 10^{10}$   & $0.74\times 10^{10}$ & $1.26\times 10^{11}$ & $1.13\times 10^{11}$ \\
   $A_{44}\ (Pa)$  & $1.29\times 10^{10}$   & $0.63\times 10^{10}$ & $0.75\times 10^{11}$ &  $0.82\times 10^{11}$\\
   $MRE=$  & 					 &				       &				   & \\
   $max\left(\frac{|\bar{\textbf{y}}_{Bi} - H\hat{\mathrm{\textbf{x}}}_{Bi}|}{\sigma_{\bar{\textbf{y}}_{Bi}}}\right)$ &  $3.74$  & $6.07$ & $5.47$ & $5.73$ \\
 \hline
  \end{tabular}
\end{center}
\end{table}

\begin{table}
\begin{center}
  \caption{Elastic constants and maximum residual error for NaCl, KCl, Cu, and CuZn using couple stress theory.}
  \vspace{10pt}
  \label{T2}
  \begin{tabular}{ccccc}
  \hline
  \textbf{} & \textbf{NaCl} & \textbf{KCl} & \textbf{Cu} & \textbf{CuZn} \\
  \hline
   $A_{11}\ (Pa)$ & $4.94\times 10^{10}$   & $4.10\times 10^{10}$ & $1.71\times 10^{11}$ &  $1.29\times 10^{11}$\\
   $A_{12}\ (Pa)$ & $1.31\times 10^{10}$   & $0.75\times 10^{10}$ & $1.27\times 10^{11}$ & $1.14\times 10^{11}$ \\
   $A_{44}\ (Pa)$  & $1.26\times 10^{10}$   & $0.59\times 10^{10}$ & $0.75\times 10^{11}$ &  $0.82\times 10^{11}$\\
   $\eta\ (N)$    & $9.68$                 & $9.17$               & $39.35 $             &  $20.90$\\
   $MRE=$  & 					 &				       &				   & \\
   $max\left(\frac{|\bar{\textbf{y}}_{Ai} - H\hat{\mathrm{\textbf{x}}}_{Ai}|}{\sigma_{\bar{\textbf{y}}_{Ai}}}\right)$ &  $2.91$  & $0.66$ & $3.96$ & $3.91$ \\
 \hline
  \end{tabular}
\end{center}
\end{table}

\begin{table}
\begin{center}
  \caption{Residual errors $\left(\frac{|\bar{\textbf{y}}_{i} - H\hat{\mathrm{\textbf{x}}}_{i}|}{\sigma_{\bar{\textbf{y}}_{i}}}\right)$ 
  for classical elasticity (model B) and couple stress theory (model A).}
  \vspace{10pt}
  \label{T}
  \begin{tabular}{cccccc}
  \hline
 $\bar{\textbf{y}}$ &&  \multicolumn{4}{c}{Residual Error}\\
  \cline{3-6}
  $\downarrow$ && \textbf{NaCl} & \textbf{KCl} & \textbf{Cu} & \textbf{CuZn} \\
  \hline
    \multirow{2}{*}{$\rho V_1^2$} & A & $2.91$   & $0.66$ & $1.47$ &  $0.85$\\
    & B & 3.73& 1.13& 1.67& 3.89\\
    \multirow{2}{*}{$2\rho V_2^2$} & A & $1.94$   & $0.21$ & $3.96$ & $3.91$ \\
    & B &0.89 & 6.07&5.47 & 4.74\\
   \multirow{2}{*}{$\rho\left(\frac{V_3+V_5}{2}\right)^2$}  & A & $0.57$   & $0.05$ & $1.34$ &  $1.60$\\
   & B & 0.26& 1.26&1.86 &1.94\\
   \multirow{2}{*}{$2\rho V_4^2$}  & A  & $1.97$                 & $0.46$               & $0.82 $             &  $0.27$\\
   & B & 3.32 & 2.48& 3.13&5.73\\
    \multirow{2}{*}{$K_{ad}$}  & 	A &	$0.19$	&	$0.08$			       &	$0.65$			   &  $0.77$\\
    & B & 1.22 & 2.12& 2.43&3.20\\
    \hline
  \end{tabular}
\end{center}
\end{table}

Besides showing the inability of classical elasticity to satisfy both wave velocities and adiabatic bulk modulus simultaneously, we will further demonstrate the performance superiority of couple stress theory over classical elasticity through information criteria. BIC and AIC of the model are given as follows \citep{Gerda2008}:
\begin{align}\label{C-S51}
BIC &= -2\ln\hat{L} + K\ln n \nonumber \\
AIC &= 2K - 2\ln\hat{L}
\end{align}
where again $K$ is the number of parameters in the statistical model, $n$ is the number of experimental measurements, and $\hat{L}$ is the maximized value of the likelihood function for the estimated model. The likelihood function for the couple stress theory model and the classical continuum mechanics model can be written, respectively, as
\begin{align}\label{C-S52}
L_A &= \frac{1}{(2\pi)^{\frac{n}{2}}|R_A|^{\frac{1}{2}}} \nonumber \\
& \times\exp \left(-\frac{1}{2}[\bar{\textbf{y}}_A - H_A{\mathrm{\textbf{x}}}_A]^TR_A^{-1}[\bar{\textbf{y}}_A - H_A{\mathrm{\textbf{x}}}_A]\right) \nonumber\\
L_B &= \frac{1}{(2\pi)^{\frac{n}{2}}|R_B|^{\frac{1}{2}}} \nonumber \\
& \times\exp \left(-\frac{1}{2}[\bar{\textbf{y}}_B - H_B{\mathrm{\textbf{x}}}_B]^TR_B^{-1}[\bar{\textbf{y}}_B - H_B{\mathrm{\textbf{x}}}_B]\right)
\end{align}
where $R_A = E[\nu_A\nu_A^T]$ and $R_B = E[\nu_B\nu_B^T]$. After maximizing Eq.~\eqref{C-S52}, we may write
\begin{align}\label{C-S53}
\hat{\mathrm{\textbf{x}}}_A &= (H_A^{T}R_A^{-1}H_A)^{-1}H_A^TR_A^{-1}\bar{\textbf{y}}_A \nonumber \\
\hat{\mathrm{\textbf{x}}}_B &= (H_B^{T}R_B^{-1}H_B)^{-1}H_B^TR_B^{-1}\bar{\textbf{y}}_B
\end{align}
Notice that Eq.~\eqref{C-S53} is the same as Eq.~\eqref{C-S28}. Here we have just showed a brief derivation of an ML solution. If we substitute Eq.~\eqref{C-S53} into Eq.~\eqref{C-S52}, we will obtain the maximized value of the maximum likelihood functions $\hat{L}_A$ and $\hat{L}_B$. Now, using the value of $\hat{L}_A$ and $\hat{L}_B$, and Eq.~\eqref{C-S51}, the values of BIC and AIC for NaCl, KCl, Cu, and CuZn are evaluated, as defined in Table~\ref{T3}.

\begin{table}
\begin{center}
  \caption{BIC and AIC values for NaCl, KCl, Cu, and CuZn.}
  \vspace{10pt}
  \label{T3}
  \begin{tabular}{ccccc}
  \hline
  \textbf{} & \textbf{NaCl} & \textbf{KCl} & \textbf{Cu} & \textbf{CuZn} \\
  \hline
   $BIC_A$ & $212.32$   & $192.93$ & $232.50$ &  $229.64$\\
   $AIC_A$ & $213.88$   & $194.49$ & $234.06$ & $231.20$ \\
  $BIC_B$  & $221.57$   & $241.03$ & $261.99$ &  $293.16$\\
   $AIC_B$ &  $222.74$  & $242.20$ & $263.16$ & $294.34$ \\
  $BIC_B - BIC_A$  & $9.25$   & $48.10$ & $29.49$ &  $63.52$\\
   $RL$ &     $1.19\times 10^{-2}$  & $4.36\times 10^{-11}$ & $4.80\times 10^{-7}$ & $1.95\times 10^{-14}$ \\
   \hline
  \end{tabular}
\end{center}
\end{table}

General rules outlined in the literature regarding BIC are as follows \citep{Gerda2008, YYang2005}: 
\begin{align}
&\bf{BIC_B - BIC_A < 2:}\nonumber \\
& \text{Weak evidence that model A is superior to model B} \nonumber \\
&\bf{2\ge BIC_B - BIC_A\le 6:}\nonumber \\
& \text{Moderate evidence that model A is superior} \nonumber \\
&\bf{6<BIC_B - BIC_A\le 10:}\nonumber \\
& \text{Strong evidence that model A is superior} \nonumber \\
&\bf{BIC_B - BIC_A > 10:}\nonumber \\
& \text{Very strong evidence that model A is superior} \nonumber
\end{align}
For AIC, the lower the relative likelihood $RL = e^{\left\{\frac{(AIC_A - AIC_B)}{2}\right\}}$ of model B, the lower the probability of model B to minimize information loss, meaning model A is superior to model B. 

The calculation of $BIC_B - BIC_A$ and RL are given in Table~\ref{T3}, and by interpretation of the general guidelines, there is no ambiguity that couple stress theory better explains linear elastic behavior for these cubic single crystals at high frequencies/small scales as compared to classical continuum mechanics theory.  More specifically, there is very strong evidence that couple stress theory is superior to classical theory for KCl, Cu and CuZn, while strong evidence is provided for the superiority for NaCl.  This slightly lesser support for the case of NaCl is consistent with the increased sensitivity of the couple stress parameter $\eta$ found for that material. 

\subsection{Material Length Scales}
Based upon the calculated adiabatic elastic constants for the four cubic crystals as summarized in Table~\ref{T2}, we may now estimate a length scale $l$ for each material.  Under consistent couple stress theory, size-dependent response becomes significant at characteristic lengths on the order of $l$ and smaller, as illustrated for several typical examples in \cite{Ali2012} and \cite{Darrall2014}.  Consequently, if $l$ for a material is on the order of the atomic spacing, then couple stress theory would have limited value, because at such scales a continuum representation would surely be in question.  Let us next estimate $l$ for the four materials under consideration here.

For isotropic materials in Sect. \ref{waveprop}, we defined $l^2$ as the ratio of the couple stress modulus $\eta$ to the shear modulus $\mu$ \citep{Ali2011}.  For cubic crystals, we may follow a similar line of reasoning to define the material length scale as
\begin{align}\label{C-S55}
l=\sqrt{\eta/A_{44}}
\end{align}
Then, from the elastic constants for consistent couple stress theory presented in Table~\ref{T2}, we find length scales of $28\mu$m, $39\mu$m, $23\mu$m, and $16\mu$m for NaCl, KCl, Cu, and CuZn, respectively.  On the other hand, the lattice constant for each of these crystals is less than $1$nm.  Thus, $l$ is orders of magnitude larger than the atomic spacing.  In fact, for a volume of size  $l^3$, there are many trillions of atoms and one would clearly expect continuum theory to apply at scales well below the material length scale $l$.  In turn, this suggests strongly that consistent couple stress theory may be quite appropriate to predict size-dependent stiffness of micro- and nano-scale components, capture stress concentration factors around small scale structural features, characterize stress fields near cracks and sharp notches, and contribute to the understanding of dislocations and disclinations.  There is a range of size-dependent multiphysics problems that are also of interest \citep{Ali2013piezo, Ali2014thermo}.  Further development of consistent couple stress theory in all of these directions, and others, would seem warranted.

\section{Conclusion}
Raman and his colleagues in the middle of the twentieth century were perhaps the first to call into question the elastic characterization of cubic crystals within the classical Cauchy theory, which requires three material constants.  In particular, the Raman group found that it is not possible to capture the wave velocities measured from ultrasonic experiments and the adiabatic bulk modulus within reasonable bounds.  Consequently, in this paper, we first cited a number of candidate size-dependent elasticity theories and 
then explored in detail the recently developed consistent couple stress theory.  This non-classical theory not only complies with wave propagation in the media (i.e., it does not lead to any spurious waves), but also has the ability to capture better linear elastic material behavior at high frequency/small scale.  Furthermore, a rigorous statistical analysis reveals strong to very strong evidence that couple stress theory is superior to classical Cauchy elasticity for representing the response for four specific single crystals with cubic structure. Additional experiments at multiple frequencies are recommended to fill in the unexplored region in the phonon dispersion relations and to test consistent couple stress theory more thoroughly.  This, in turn, can lead to a better fundamental understanding of material response and toward the development of a predictive framework for engineering applications.  This refers, of course, not only to cubic crystals, but for the broad range of materials finding use in modern micro- and nano-scale technologies.

\section{Acknowledgment}
The authors wish to acknowledge the contribution of Professor Puneet Singla, who recommended information theoretic approaches to compare the couple stress and classical elastic formulations.



\baselineskip 15pt

\bibliographystyle{apa}
\bibliographystyle{elsarticle-num}
\bibliography{main}

\begin{thebibliography}{}

\bibitem[\protect\astroncite{Altan and Aifantes}{1997}]{AltanAifantes1997}
Altan, B. and Aifantes, E. (1997).
\newblock On some aspects in the special theory of gradient elasticity.
\newblock {\em Journal of the Mechanical Behavior of Materials}, 8:231--282.

\bibitem[\protect\astroncite{Askes and Aifantes}{2011}]{AskesAifantes2011}
Askes, H. and Aifantes, E. (2011).
\newblock Gradient elasticity in statics and dynamics: An overview of
  formulations, length scale identification procedures, finite element
  implementations and new results.
\newblock {\em International Journal of Solids and Structures}, 48:1962--1990.

\bibitem[\protect\astroncite{Chen and Lee}{2003}]{Youping1}
Chen, Y. and Lee, J. (2003).
\newblock Determining material constants in micromorphic theory through phonon
  dispersion relations.
\newblock {\em International Journal of Engineering Science}, 41:871--886.

\bibitem[\protect\astroncite{Chen et~al.}{2003}]{Youping2}
Chen, Y., Lee, J., and Eskandarian, A. (2003).
\newblock Examining the physical foundation of continuum theories from the
  viewpoint of phonon dispersion relation.
\newblock {\em International Journal of Engineering Science}, 41:61--83.

\bibitem[\protect\astroncite{Claeskens and Hjort}{2008}]{Gerda2008}
Claeskens, G. and Hjort, N. (2008).
\newblock {\em Model Selection and Model Averaging}.
\newblock Cambridge Series in Statistical and Probabilistic Mathematics.
  Cambridge University Press, 1st edition.

\bibitem[\protect\astroncite{Cosserat and Cosserat}{1909}]{Cosserat1909}
Cosserat, E. and Cosserat, F. (1909).
\newblock {\em Theorie des Corps Deformables}.
\newblock Hermann et Fils, Paris.

\bibitem[\protect\astroncite{Cowin and Nunziato}{1983}]{Cowin1983}
Cowin, S. and Nunziato, J. (1983).
\newblock Linear elastic materials with voids.
\newblock {\em J. Elasticity}, 13:125--147.

\bibitem[\protect\astroncite{Crassidis and Junkins}{2012}]{Crassidis2012}
Crassidis, J. and Junkins, J. (2012).
\newblock {\em Optimal Estimation of Dynamic Systems}.
\newblock CRC Press, Taylor and Francis Group, LLC, 2nd edition.

\bibitem[\protect\astroncite{Darrall et~al.}{2014}]{Darrall2014}
Darrall, B., Dargush, G., and Hadjesfandiari, A. (2014).
\newblock Finite element lagrange multiplier formulation for size-dependent
  skew-symmetric couple-stress planar elasticity.
\newblock {\em Acta Mechanica}, 225:195--212.

\bibitem[\protect\astroncite{DiVincenzo}{1986}]{DiVincenzo1986}
DiVincenzo, D.~P. (1986).
\newblock Dispersive corrections to continuum elastic theory in cubic crystals.
\newblock {\em Physical Review B}, 34:3450--3465.

\bibitem[\protect\astroncite{Eringen}{1968a}]{Eringen1968-1}
Eringen, A. (1968a).
\newblock {\em Mechanics of micromorphic continua}.
\newblock IUTAM symposium, Mechanics of Generalized Continua. Springer Verlag.

\bibitem[\protect\astroncite{Eringen}{1968b}]{Eringen1968}
Eringen, A. (1968b).
\newblock {\em Theory of micropolar elasticity}, volume~2 of {\em In Fracture}.
\newblock Academic Press.

\bibitem[\protect\astroncite{Eringen}{1972}]{Eringen1972}
Eringen, A. (1972).
\newblock Linear theory of nonlocal elasticity and dispersion of plane waves.
\newblock {\em Int. J. Engng Sci}, 10:425--435.

\bibitem[\protect\astroncite{Every}{}]{Every2005}
Every, A.
\newblock {\em Physical Review B}, page 104302.

\bibitem[\protect\astroncite{Every et~al.}{}]{Everyetal2006}
Every, A., Kaplunov, J., and Rogerson, G.
\newblock {\em Physical Review B}, page 184307.

\bibitem[\protect\astroncite{Galt}{1948}]{Galt1948}
Galt, J. (1948).
\newblock Mechanical properties of nacl, kbr, kcl.
\newblock {\em Physical Review}, 73(12):1460.

\bibitem[\protect\astroncite{Good}{1941}]{Good1941}
Good, W. (1941).
\newblock Rigidity modulus of beta-brass single crystals.
\newblock {\em Physical Review}, 60:605.

\bibitem[\protect\astroncite{Hadjesfandiari}{2013}]{Ali2013piezo}
Hadjesfandiari, A. (2013).
\newblock Size-dependent piezoelectricity.
\newblock {\em International Journal of Solids and Structures}, 50:2781--2791.

\bibitem[\protect\astroncite{Hadjesfandiari}{2014}]{Ali2014thermo}
Hadjesfandiari, A. (2014).
\newblock Size-dependent thermoelasticity.
\newblock {\em Latin American Journal of Solids and Structures}, 11:1679--1708.

\bibitem[\protect\astroncite{Hadjesfandiari and Dargush}{2011}]{Ali2011}
Hadjesfandiari, A. and Dargush, G. (2011).
\newblock Couple stress theory for solids.
\newblock {\em International Journal of Solids and Structures}, 48:2496--2510.

\bibitem[\protect\astroncite{Hadjesfandiari and Dargush}{2012}]{Ali2012}
Hadjesfandiari, A. and Dargush, G. (2012).
\newblock Boundary element formulation for plane problems in couple stress
  elasticity.
\newblock {\em International Journal for Numerical Methods in Engineering},
  89:618--636.

\bibitem[\protect\astroncite{Hadjesfandiari and Dargush}{2015}]{Ali-Gary2015}
Hadjesfandiari, A. and Dargush, G. (2015).
\newblock Foundations of couple stress theory.
\newblock arXiv 1509-06299.

\bibitem[\protect\astroncite{Huntington}{1947}]{Huntington1947}
Huntington, H. (1947).
\newblock Ultrasonic measurements on single crystals.
\newblock {\em Physical Review}, 72(4):321.

\bibitem[\protect\astroncite{Koiter}{1964}]{Koiter1964}
Koiter, W. (1964).
\newblock Couple stresses in the theory of elasticity.
\newblock {\em Proc. Ned. Akad. Wet. (B)}, 67:17--44.

\bibitem[\protect\astroncite{Lakes}{1995}]{Lakes1995}
Lakes, R. (1995).
\newblock {\em Experimental methods for study of Cosserat elastic solids and
  other generalized elastic continua. \text{In:} Continuum models for materials
  with micro-structure, ed. H. Muhlhaus, Ch. 1, p.1-22}.
\newblock J. Wiley, New York.

\bibitem[\protect\astroncite{Lazar et~al.}{2005}]{Lazar2005}
Lazar, M., Maugin, G., and Aifantis, E. (2005).
\newblock On dislocations in a special class of generalized elasticity.
\newblock {\em Phys. Status Solidi B}, 242:2365--2390.

\bibitem[\protect\astroncite{Lazarus}{1949}]{Lazarus1949}
Lazarus, D. (1949).
\newblock The variation of the adiabatic elastic constants of kcl, nacl, cuzn,
  cu, and al with pressure to 10,000 bars.
\newblock {\em Physical Review}, 76(4):545.

\bibitem[\protect\astroncite{Maranganti and Sharma}{2007a}]{Maranganti2007}
Maranganti, R. and Sharma, P. (2007a).
\newblock Length scales at which classical elasticity breaks down for various
  materials.
\newblock {\em Physical Review Letters}, 98:195504.

\bibitem[\protect\astroncite{Maranganti and Sharma}{2007b}]{Maranganti2007jmps}
Maranganti, R. and Sharma, P. (2007b).
\newblock A novel atomistic approach to determine strain-gradient elasticity
  constants: Tabulation and comparison for various metals, semiconductors,
  silica, polymers and the (ir) relevance for nanotechnologies.
\newblock {\em Journal of the Mechanics and Physics of Solids}, 55:1823--1852.

\bibitem[\protect\astroncite{Mindlin}{1964}]{Mindlin1964}
Mindlin, R. (1964).
\newblock Micro-structure in linear elasticity.
\newblock {\em Arch. Rational Mech. Analysis}, 16:51--78.

\bibitem[\protect\astroncite{Mindlin}{1965}]{Mindlin1965}
Mindlin, R. (1965).
\newblock Stress functions for a cosserat continuum.
\newblock {\em International Journal of Solids and Structures}, 1:265--271.

\bibitem[\protect\astroncite{Mindlin and Tiersten}{1962}]{Mindlin1962}
Mindlin, R. and Tiersten, H. (1962).
\newblock Effect of couple-stresses in linear elasticity.
\newblock {\em Arch. Rational Mech. Analysis}, 11:415--488.

\bibitem[\protect\astroncite{Nowacki}{1986}]{Nowacki1986}
Nowacki, W. (1986).
\newblock {\em Theory of Asymmetric Elasticity}.
\newblock Pergamon Press, Oxford.

\bibitem[\protect\astroncite{Pandya et~al.}{2001}]{Platinum}
Pandya, C., Vyas, P., Pandya, T., Rani, N., and Gohel, V. (2001).
\newblock An improved lattice mechanical model for fcc transition metals.
\newblock {\em Physica B}, 307:138--149.

\bibitem[\protect\astroncite{Puri and Cowin}{1985}]{Puri-Cowin1985}
Puri, P. and Cowin, S. (1985).
\newblock Plane waves in linear elastic materials with voids.
\newblock {\em Journal of Elasticity}, 15(2):167--183.

\bibitem[\protect\astroncite{Raman and
  Krishnamurti}{1955}]{Raman-Krishna9-1955}
Raman, C. and Krishnamurti, D. (1955).
\newblock Evaluation of the four elastic constants of some cubic crystals.
\newblock {\em Proceedings of the Indian Academy of Sciences - Section A},
  42(3):111--130.

\bibitem[\protect\astroncite{Raman and Viswanathan}{1955}]{Raman-Viswa8-1955}
Raman, C. and Viswanathan, K. (1955).
\newblock On the theory of the elasticity of crystals.
\newblock {\em Proceedings of the Indian Academy of Sciences - Section A},
  42(2):51--70.

\bibitem[\protect\astroncite{Shodja et~al.}{2013}]{Shodjaetal2013}
Shodja, H., Zaheri, A., and Tehranchi, A. (2013).
\newblock Ab initio calculations of characteristic lengths of crystalline
  materials in first strain gradient elasticity.
\newblock {\em Mechanics of Materials}, 61:73--78.

\bibitem[\protect\astroncite{Viswanathan}{1955}]{Viswa1955}
Viswanathan, K. (1955).
\newblock The theory of the elasticity of crystals.
\newblock {\em Proceedings of the Indian Academy of Sciences - Section A},
  41(3):98--116.

\bibitem[\protect\astroncite{Voigt}{1887}]{Voigt1887}
Voigt, W. (1887).
\newblock {\em Theoretische Studien uber die Elasticitatsverhaltnisse der
  Krystalle}.
\newblock Number~34. Abh. Ges. Wiss. Gottingen.

\bibitem[\protect\astroncite{Yang et~al.}{2002}]{Yang2002}
Yang, F., Chong, A., Lam, D., and Tong, P. (2002).
\newblock Couple stress based strain gradient theory for elasticity.
\newblock {\em International Journal of Solids and Structures},
  47(1):2731--2743.

\bibitem[\protect\astroncite{Yang}{2005}]{YYang2005}
Yang, Y. (2005).
\newblock Can the strengths of aic and bic be shared? a conflict between model
  indentification and regression estimation.
\newblock {\em Biometrika}, 92(4):937--950.

\end{thebibliography}







\end{document}